\documentclass[fleqn]{2020SCGE}

\newcommand{\araa}{Annu.~Rev.~Astron.~Astrophys.~}

\newcommand{\apj}{Astrophys.~J.~}
\newcommand{\apjl}{Astrophys.~J.~Lett.~}

\newcommand{\mnras}{Mon.~Not.~R.~Astron.~Soc.~}
\newcommand{\aap}{Astron.~Astrophys.~}
\newcommand{\aj}{Astron.~J.~}

\newcommand{\pasp}{Publ.~Astron.~Soc.~Pac.~}

\newcommand{\nat}{Nature~}

\newcommand{\jcap}{J.~Cosmol.~Astropart.~Phys.~}

\usepackage{hyperref} 
\usepackage{subfigure}
\usepackage{amsmath}
\usepackage{cases}
\usepackage{tabularx}
\usepackage{multirow, booktabs}
\usepackage{subfigure}
\usepackage{xcolor}
\usepackage{graphicx}
\usepackage{threeparttable}

\begin{document}

\ensubject{subject}

\ArticleType{Article}
\SpecialTopic{SPECIAL TOPIC: }
\Year{2024}
\Month{}
\Vol{?}
\No{?}
\DOI{??}
\ArtNo{000000}
\ReceiveDate{XXX}
\AcceptDate{YYY}


\title{Forecasting Supernova Observations with the CSST: I. Photometric Samples}{Forecasting Supernova Observations with the CSST: I. Photometric Samples}

\author[1,2,3]{Chengqi Liu}{}%
\author[3]{Youhua Xu}{}
\author[3]{Xianmin Meng}{}
\author[3]{Xin Zhang}{}
\author[4]{Shi-Yu Li}{}
\author[5,6]{Yuming Fu}{}
\author[7]{\\Xiaofeng Wang}{}
\author[3,8]{Shufei Liu}{}
\author[3,8]{Zun Luo}{}
\author[3,8]{Guanghuan Wang}{}
\author[3,2]{Hu Zhan}{zhanhu@nao.cas.cn}%

\AuthorMark{Chengqi Liu}

\AuthorCitation{Chengqi Liu, Youhua Xu, Xianmin Meng, et al}

\address[1]{Department of Astronomy, School of Physics, Peking University, Beijing 100871, PR China}
\address[2]{Kavli Institute for Astronomy and Astrophysics, 
Peking University, Beijing 100871, PR China}
\address[3]{Key Laboratory of Space Astronomy and Technology, National Astronomical Observatories, Chinese Academy of Sciences, Beijing, 100101, China}
\address[4]{Beijing Planetarium, Beijing Academy of Science and Technology, Beijing 100044, PR China}
\address[5]{Leiden Observatory, Leiden University, P.O. Box 9513, NL-2300 RA Leiden, The Netherlands}
\address[6]{Kapteyn Astronomical Institute, University of Groningen, P.O. Box 800, NL-9700 AV Groningen, The Netherlands}
\address[7]{Physics Department, Tsinghua University, Beijing, 100084, China}
\address[8]{School of Astronomy and Space Science, University of Chinese
Academy of Sciences, Beijing 100049, PR China}


\abstract{The China Space Station Telescope (CSST, also known as Xuntian) is a serviceable two-meter-aperture wide-field telescope operating in the same orbit as the China Space Station. 
The CSST plans to survey a sky area of 17,500~deg$^2$ of the medium-to-high Galactic latitude to a depth of 25--26~AB~mag in at least 6 photometric bands over 255--1000~nm. Within such a large sky area, slitless spectra will also be taken over the same wavelength range as the imaging survey.
Even though the CSST survey is not dedicated to time-domain studies, it would still detect a large number of transients, such as supernovae (SNe).
In this paper, we simulate photometric SN observations based on a strawman survey plan using the \textsc{Sncosmo} package. 
During its 10-year survey, the CSST is expected to observe about 5
million SNe of various types.
With quality cuts, we obtain a ``gold'' sample that comprises roughly 7,400 SNe Ia, 2,200 SNe Ibc, and 6,500 SNe II candidates with correctly classified percentages reaching 91\%, 63\%, and 93\% (formally defined as classification precision),  respectively. 
The same survey can also trigger alerts for the detection of about 15,500 SNe Ia (precision 61\%)
and 2,100 SNe II (precision 49\%) candidates at least two days before the light maxima.
Moreover, the near-ultraviolet observations of the CSST will be able to catch hundreds of shock-cooling events serendipitously every year.
These results demonstrate that the CSST can make a potentially significant
contribution to SN studies.}



\maketitle


\begin{multicols}{2}
\section{Introduction} \label{sec:intro}
Observations of SNe play a crucial role in measuring the accelerated
expansion of the Universe and studying the SNe themselves.
Over the past two decades, 
projects such as the \Authorfootnote 
ESSENCE supernova Survey \cite{2007ApJ...666..674M},
the Catalina Real-Time 
Transient Survey  \cite{2011arXiv1102.5004D}, 
the Dark Energy Survey \cite{2016MNRAS.460.1270D}, 
the Panoramic Survey Telescope and Rapid 
Response System \cite{2016arXiv161205560C},
the All-Sky Automated Survey for Supernovae
\cite{2017PASP..129j4502K},
and the Zwicky Transient Facility  \cite{2019PASP..131a8002B}
acquired a large number of high-quality SNe light curves (LCs) and spectra.
They have shed light on many questions in SN science and cosmology \cite{2007ApJ...666..694W,2009ApJ...696..870D,2016Sci...351..257D,2018ApJ...857...51J,2019ApJ...872L..30A,2020ApJ...904...35P}, 
but there is always more to explore.  
For instance, the progenitor stars and the mechanisms of SN explosions remain unresolved issues.
The need for more extensive samples of SNe to tackle these issues has been a key driver for several new survey projects such as the Wide Field Survey Telescope (WFST) in Lenghu, China \cite{2023SCPMA..6609512W}
and the Vera C. Rubin Observatory (Rubin) on Cerro Pach\'on in Chile \cite{lsstsciencecollaboration2009lsst,2019ApJ...873..111I,2018RPPh...81f6901Z}.

The upcoming 2-meter China Space Station Telescope 
\cite{2011SSPMA..41.1441Z,Zhan2021} plans to start a large-scale multiband imaging and slitless spectroscopy survey around 2026. 
The survey, taking approximately 70\% time of the 10-year mission, will cover an area of about 17,500 $\rm deg^2$ in 7 photometric bands and 3 spectroscopic bands. 
While the CSST places an emphasis on cosmology and extragalactic science, the survey can be quite useful for time-domain studies as well because of its depth, image quality, and near-ultraviolet (NUV) capability.  

The single-exposure depth of the CSST wide survey will be about 25-26~mag on average, enabling discoveries of SNe at early stages or high redshifts. 
The size of the CSST point spread function, quantified by the radius encircling 80\% energy, is specified to be no more than 0.15$^{\prime\prime}$. The combination of high-quality data from space-based observations with those from ground-based telescopes will enable detailed studies on SN progenitors, characteristics of host galaxies, and statistics of SN locations within the host galaxies \cite{2002MNRAS.336L..17F,2016AJ....152..154G,2017hsn..book..693V}.
The NUV band of the CSST allows it to catch shock-cooling events, which can further deepen our understanding of the SN explosion mechanism \cite{2009AJ....137.4517B,2010ApJ...721.1627M,2017ApJ...838..130S}.

The CSST is expected to 
detect a large number of SNe and make potentially important contributions to SN research.
Previous studies \cite{2023SCPMA..6629511L,2024MNRAS.530.4288W} have
explored SN Ia detection
and cosmological constraints with a proposed 9~deg$^2$ CSST ultra-deep field.
In this paper, we aim to make a realistic assessment of the CSST's capability to detect SNe
in its wide survey and deep fields using pointings generated by simulated operations.

The paper is structured as follows. Section \ref{sec:mock} describes the simulation of the CSST SN observations and the classification methods for the mock data. Section \ref{sec:samples} presents the resulting SN samples. 
Discussions and conclusions are given in Section \ref{sec:summary}. 
Throughout this paper, we adopt a flat $\Lambda$CDM model with 
parameters $H_0=70~\rm km~s^{-1}~Mpc^{-1}$, $\Omega_m=0.3$, and $\Omega_{\Lambda}=0.7$. 
Unless otherwise specified, the default magnitude system utilized is the AB magnitude system. 

\begin{table*}[t]
\centering
\footnotesize
\caption{Single-exposure imaging depth of the CSST survey}
\label{table:multi-color imaging magnitude}
\begin{tabular*}{\textwidth}{@{\extracolsep{\fill}}cccccccccc} 
    \hline \hline
        Component & Survey area $\rm (deg^2)$ & Exposure time 
        (s) & \multicolumn{7}{c}{Limiting magnitude (point source,
        5 $\sigma$)} \\
		\hline
         & & & NUV & u & g & r & i & z & y \\
        \cmidrule(r){4-10}
         Wide & 17,500 & 150 & 24.53 & 24.90 & 25.76 & 25.48 & 25.32 & 24.77 & 23.56 \\

         Deep & 400 & 250 & 25.05 & 25.41 & 26.17 & 25.88 & 25.72 & 25.20 & 24.05 \\
        \hline
\end{tabular*}
\end{table*}

\section{Mock observations of supernova} \label{sec:mock}
\subsection{Survey specifications}

\begin{figure}[H]
\centering
\includegraphics[scale=0.5,trim=170 105 10 90,clip]{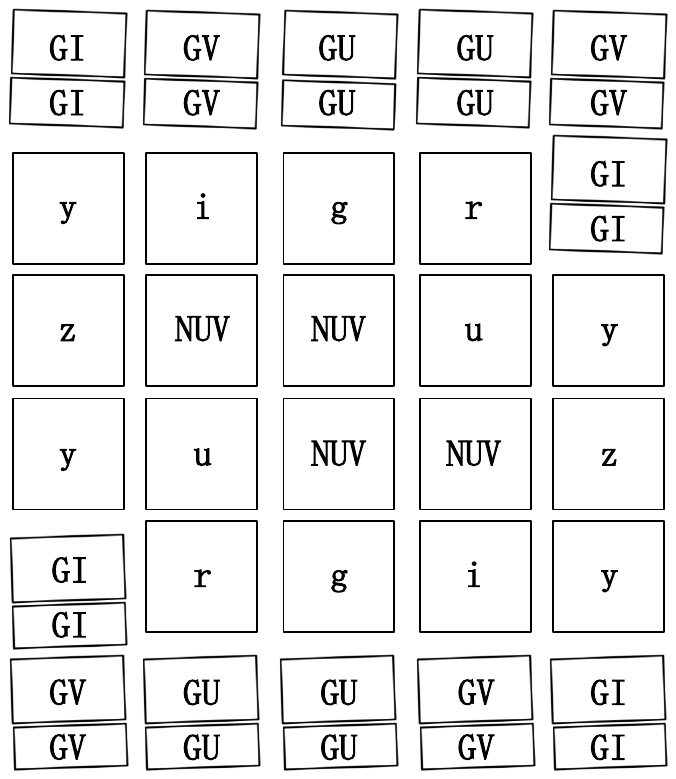}
\caption{
    The arrangement of the survey camera's main focal plane. The FoV of the photo-sensitive area is about 1.07~$\rm deg^2$ total. 
    There are seven filter bands (NUV, u, g, r, i, z, and y) and three grating bands (GU, GV, and GI). Each band covers two or four detectors. The dispersion direction of the gratings on the top and that of the gratings on the bottom are at a small angle of $2^\circ$.
    }
\label{fig:focal planes}
\end{figure}

\begin{figure}[H]
\centering
\includegraphics[scale=0.6]{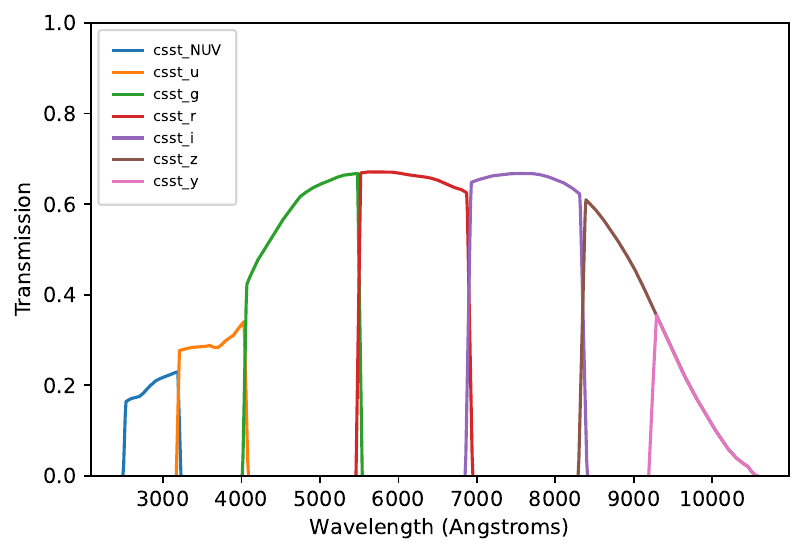}
\caption{
    System throughputs of the CSST photometric bands. 
    }
\label{fig:throughput curve}
\end{figure}

The CSST adopts a Cook-type off-axis three-mirror anastigmat system, 
which achieves high image quality within a large field of view (FoV).
Fig.~\ref{fig:focal planes} illustrates the arrangement of filters and gratings of the CSST survey camera. The focal plane consists of 30 detectors, each with a filter or two gratings mounted atop.
In this paper, we only consider the case of photometric observations.
Fig.~\ref{fig:throughput curve} shows the system throughputs of the CSST photometric bands from 255~nm to 1000~nm. 
The survey comprises both a wide component and a deep component, with stacked depths of $g\sim 26.3$ and 27.5, respectively \cite{Zhan2021}.
An ultra-deep field of $9\deg^2$ reaching $g\sim 28$ is also under consideration.
The single-exposure depths of the CSST imaging survey are given in Table~\ref{table:multi-color imaging magnitude}. 
Details of the wide field survey and the deep field survey are outlined below.

\textit{Wide Field Survey}:
The wide field survey primarily covers Galactic latitudes $|b| \ge 15^{\circ}$ and ecliptic latitudes $|\beta|\ge 15^{\circ}$, spanning an area of about 17,500 $\rm deg^2$.
Each detector covers the entire area once with a nominal exposure time of 150 seconds. As a result, each patch of the sky receives 2 observations in the u, g, r, i, and z bands and 4 observations in the NUV and y bands. This leads to a total number of 18 photometric observations in each sky patch of the wide survey.

\textit{Deep Field Survey}:
The deep field survey covers a sky area of 400 $\rm deg^2$, whose selection has not been finalized yet. 
In the current simulation of the CSST operations, 8 fields are selected for demonstration purposes. 
Each detector covers all the deep fields 4 times with a nominal exposure time of 250 seconds per visit, resulting in a total of 72 visits summed over all photometric bands. 

Over the 10-year survey operations,  
the CSST is expected to take about 650,000 exposures, including about 60,000 exposures in the deep fields.
Using Healpix \cite{2005ApJ...622..759G}, 
we evenly divide the entire 17,500 $\rm deg^2$ survey 
area into about 1,500,000 small sky patches, with each patch covering an area of about 0.013 $\rm deg^2$. 
Subsequently, we extract the observation time series for each sky patch from the pointing sequence generated by the simulation of operations. 
The median interval between two consecutive visits, regardless of which two bands are observed, is around six weeks in the wide survey and two weeks in the deep survey. 
About 28\% consecutive visits of the same band are made within one day, in which case only one visit is counted in our analyses.

\subsection{Supernova simulations} \label{SN simulations}
We use the Python package \textsc{Sncosmo} to simulate SN observations based on the CSST survey specifications. 
To generate realistic SN LCs, we need to take into consideration the following factors: the SN volumetric rate, SN models, extinction correction, and signal-to-noise ratio calculation.

\subsubsection{Supernova volumetric rate} \label{SNe rate}
The volumetric rate of SNe is the number of SNe within a given timespan and a fixed co-moving volume, which can be described as a function of redshift $z$.
In our simulations, the maximum redshift of the SNe is set to 1.4 based on the single-exposure detection limit of the CSST.
We adopt a power-law model to describe the event rate of SN Ia  \cite{2008ApJ...682..262D,2018ApJ...867...23H}
\begin{eqnarray} \label{eq:Ia rate}
R_{Ia}(z) = \begin{cases}
2.5\times (1+z)^{1.5} & z\le 1, \\
9.7\times (1+z)^{-0.5} & 1<z<3,
\end{cases}
\end{eqnarray}
and that of Core-collapse supernova (CCSN) \cite{2009A&A...499..653B,lsstsciencecollaboration2009lsst} 
\begin{eqnarray}\label{eq:Cc rate}
R_{CC}(z) = 6.8\times(1+z)^{3.6}.
\end{eqnarray}
The unit of the rates is 
$10^{-5}~\rm h_{70}^{3}~Mpc^{-3}~yr^{-1}$.

\subsubsection{Supernova models}
SALT2 \cite{2007A&A...466...11G,2010A&A...523A...7G} 
is an empirical model depicting the spectro-photometric evolution of SNe Ia 
over time. It utilizes an extensive dataset comprising templates derived from 
LCs and spectra of both nearby and distant SNe Ia. 
SALT2 provides the average spectral sequence of SNe Ia and identifies their 
principal variability components, including a color variation law.

We adopt the SALT2-extended model \cite{2018PASP..130k4504P}
which provides a wider rest-frame wavelength coverage than the SALT2 model.
The SALT2-extended model allows 
us to measure the distance moduli in the spectral wavelength range of 30 nm to 1800 nm, which is essential for generating LCs in all the CSST photometric bands.
The model flux density of a SN Ia at a rest-frame wavelength $\lambda$ can be expressed as
\begin{multline}\label{eq:Ia temp}
F(p,\lambda)= \\
x_0\times[M_0(p,\lambda)+x_1M_1(p,\lambda)]\times \exp(c'CL(\lambda)),
\end{multline}
where $p$ represents the rest-frame time since the date of maximum luminosity 
in the \textit{B} band (referred to as the phase), $M_0(p,\lambda)$ denotes 
the average spectral sequence, $M_1(p,\lambda)$
represents the first-order variation, and $CL(\lambda)$ represents the average 
color correction law. The parameters $x_0,~ x_1 $, and $c'$ are the amplitude, stretch, and color of the LC, respectively. 

CCSNe exhibit a more heterogeneous nature compared to SNe Ia. Unlike SNe Ia, there is currently no parameterized model available to describe the diversity observed in the LCs of CCSNe. Therefore, we use time series template models to simulate CCSN observations.
The spectral flux density of the time series template model is given by
\begin{eqnarray}\label{eq:Cc temp}
 F(p,\lambda) = A \times M(p,\lambda),
\end{eqnarray}
where $M(p,\lambda)$ is the relative flux at a phase $p$ and a wavelength $\lambda$. $A$ (amplitude) is the single free parameter of the model.

Two sets of CCSN template models can effectively depict the spectral evolution sequence 
of CCSN over time.
The first set comprises 40 templates from the Supernova Photometric Classification Challenge \cite{2010PASP..122.1415K}.
The second set contains composite spectral templates compiled from the literature, which are known as the Nugent templates \cite{2002PASP..114..803N}.
The wavelength range of Nugent templates spans from 100 nm to 1500 nm, 
providing sufficient coverage.
We adopt the Nugent templates as the default templates 
to simulate CCSNe due to their compatibility with the CSST filters and their simplicity.

\begin{figure*}[t]
\centering
\begin{subfigure}{}
\includegraphics[width=2\columnwidth]{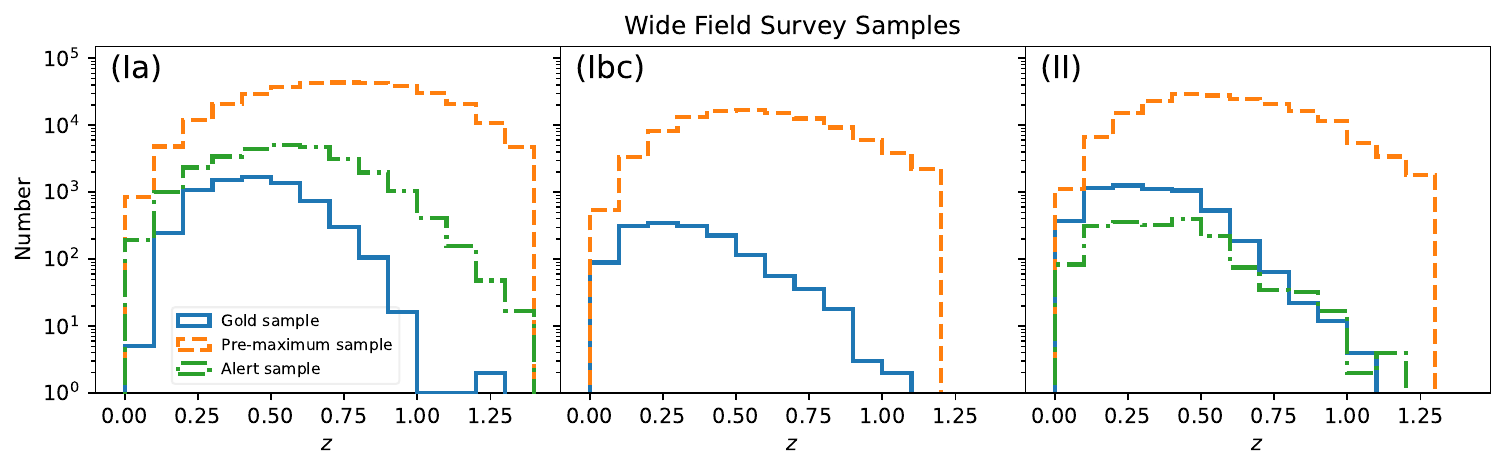} 
\end{subfigure}
\begin{subfigure}{}
\includegraphics[width=2\columnwidth]{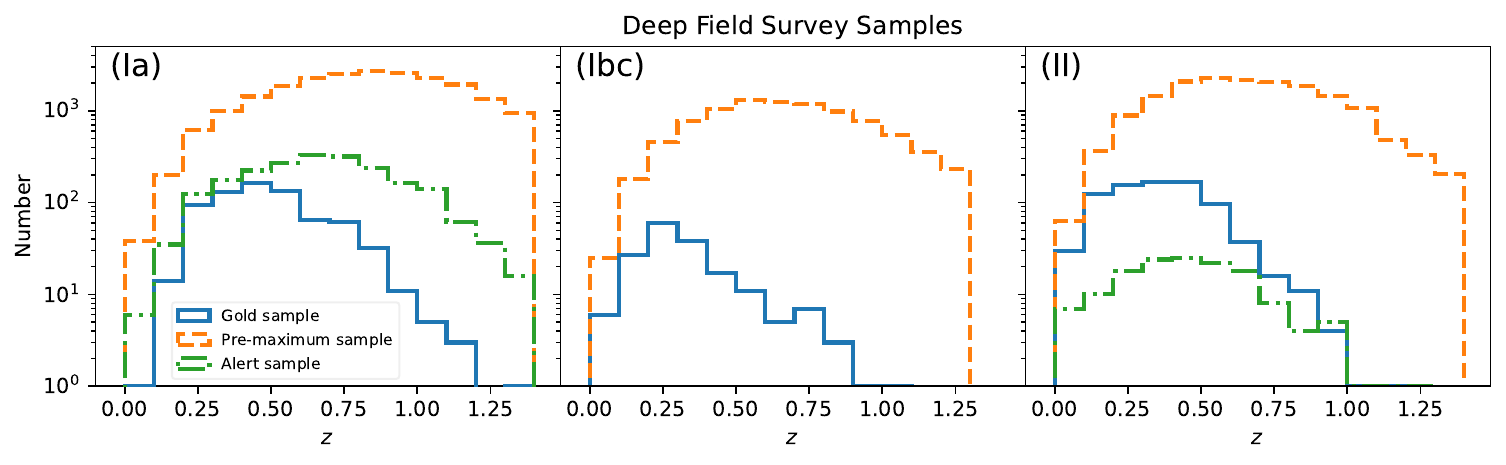}
\end{subfigure}

\caption{
Redshift distributions of  
SN samples in the wide field survey 
(\textit{upper panels}) and the deep field survey 
(\textit{lower panels}). 
The blue, orange, and green represent 
the gold sample,
pre-maximum sample, and alert sample, respectively. 
The SN types are given by the input catalog, not by classification.
}
\label{fig:redshift distribution of CSST SN}
\end{figure*}

\subsubsection{Extinction law}
Dust in the MW and host galaxy will affect the 
shape of an observed SN spectrum, which in turn affects the LCs.
We adopt the F99 dust extinction law with $\rm R_V=3.1$ \cite{1999PASP..111...63F} for both the MW and the host galaxy, with the dust extinction parameters (color excesses) being $mwebv$ and $hostebv$, respectively.
We query the SFD dust map \cite{1998ApJ...500..525S} for $mwebv$ and $hostebv$ based on the SN location with the Python package \textsc{sfdmap}.
In our simulation, the $mwebv$ and $hostebv$ parameters
have a mean value of 0.04 mag.

\subsubsection{Signal-to-noise ratio calculation}
The signal-to-noise ratio (SNR) is given by
\begin{eqnarray}\label{eq:SNR}
\frac{S}{N} = \frac{N_{obj}}{\sqrt{N_{obj}+n_{pix}(N_{sky}+N_{D}+N_{R}^2+N_{other}^2)}},
\end{eqnarray}
where $n_{pix}$ is the number 
of pixels involved in the SNR calculation, $N_{obj}$ represents the 
number of photo-electrons collected from the SN, $N_{sky}$ and  $N_D$ are the contribution per pixel from the sky background and dark current, respectively, $N_R$ denotes the readout noise, and $N_{other}$ is manually set to take into account other noises that leave a margin of 0.3~mag in the limiting magnitude.

The number of photo-electrons collected
from a target can be described by
\begin{eqnarray}\label{eq:Nobj}
N_{obj} = \int \frac{f_{\lambda}(m_{AB})}{hc/\lambda} \cdot 
\tau(\lambda) \cdot d\lambda \cdot Area \cdot t \cdot 80\%,
\end{eqnarray}
where $f(\lambda)$ is the spectral 
flux of the target, $h$ is Planck's constant, 
$c$ is the speed of light, $Area$ is the area of the primary mirror of the telescope,
$\tau(\lambda)$ is the system throughput in Fig.~\ref{fig:throughput curve}, and $t$ is the exposure time. We assume 80\% of the photons from the SN is incident on the central $n_{pix}$ pixels in the calculation.

With the angular size of a single pixel being $0.074^{\prime \prime}$, $n_{pix}$ is then $\pi(0.15/0.074)^2 \approx 13$.
The average dark current is 0.02 $\rm e^-/s/pix$, 
the readout noise is 5 $\rm e^-/pix$, and the 
readout time is about 40 seconds. 
The sky background levels scaled from the Hubble Space Telescope observations \cite{2023acsi.book...23R} are 0.0026, 0.018, 0.16, 0.21, 0.21, 0.13, and 0.038 $\rm e^-/s/pix$ for the seven photometric bands of the CSST, respectively.

\begin{table}[H]
\footnotesize
\caption{The \textit{B}-band peak absolute AB magnitude, dispersion, 
and fraction of different Core-collapse supernovae adopted from \cite{2015ApJ...813...93S}}
\label{table:subtypes of Cc SNe}
\begin{tabular*}{\columnwidth}{@{\extracolsep{\fill}}cccc} 
\hline \hline
Type & Peak $M_{B}$ & Dispersion & Fraction \\
\hline
Ibc & -17.10 & 0.99 & 25\% \\

IIP & -16.80 & 0.97 & 55\% \\

IIL & -17.98 & 0.90 & 20\% \\
\hline
\end{tabular*}
\end{table}

\subsubsection{Supernova mock observations}
We use the built-in models of \textsc{Sncosmo}  
to generate the SN LCs.
The SALT2-extended model includes five free parameters: the time of the \textit{B}-band peak magnitude ($t_0$), 
the redshift ($z$), the amplitude parameter ($x_0$), the stretch parameter ($x_1$), and the color parameter ($c'$).
The peak time $t_0$ is randomly assigned within the 10-year survey duration, while the redshift distributions follow those outlined in Section \ref{SNe rate}.
The \textit{B}-band peak absolute magnitudes of SNe Ia follow a normal distribution of $M_B \sim \mathcal{N}$(-19.3, 0.3) 
according to \cite{2014AJ....147..118R}, which determines the parameter $x_0$. 
The other parameters are set as follows: 
$x_1 \sim \mathcal{N}$(0, 1) and $c' \sim \mathcal{N}$(0, 0.1) 
\cite{2011A&A...534A..43B, 2013ApJ...763...88C, 2018MNRAS.477.4142D}. 

The LCs of SNe Ibc, SNe IIP, and SNe IIL are generated 
using Nugent-snIbc \cite{2005ApJ...624..880L}, 
Nugent-sn2p, and Nugent-sn2l \cite{1999ApJ...521...30G} templates, respectively.
The Nugent templates are described by 
three free parameters: $t_0$, $z$, and $amplitude$. 
The parameters $t_0$ and $z$ are set 
in the same way as that of SNe Ia.
According to \cite{2015ApJ...813...93S}, 
the average \textit{B}-band peak absolute magnitudes, 
dispersions, and relative fractions 
of different CCSNe are listed in 
Table \ref{table:subtypes of Cc SNe}.
The \textit{B}-band peak absolute magnitudes, which determine
the $amplitude$ for SNe Ibc, SNe IIP, and SNe IIL, follow normal distributions of $M_B \sim \mathcal{N}$(-17.1, 0.99), $M_B \sim \mathcal{N}$(-16.8, 0.97), 
and $M_B \sim \mathcal{N}$(-17.98, 0.90), respectively. 

For each SN in each exposure, we generate its flux with scatter incorporated in the observed band based on its LC phase and the CSST survey specifications. 
Host galaxy contamination to the SN photometry is assumed to be sufficiently low and is therefore not included in our current simulations. 
A separate study will be needed to address such contamination for the CSST, likely in combination with external data.

In total, about 5 million SNe of various types would be observed at least once with $\mathrm{SNR}\ge 5$ and hence be cataloged as real objects. 
This includes about 1,700,000 SNe Ia, 590,000 SNe Ibc, and 2,000,000 SNe II in the wide field, and about 73,000 SNe Ia, 32,000 SNe Ibc, and 110,000 SNe II in the deep fields.
Unfortunately, most of these SNe cannot be correctly identified as SNe with CSST data alone.

\subsection{Supernova classification}
There are mainly two types of SN photometric classification methods: machine learning (ML) and empirical approaches. ML typically 
requires a large number of well-observed training samples and a 
sufficient number of observation points to identify candidate SNe \cite{2013MNRAS.435.1047B,2016MNRAS.457.3119D,2016JCAP...12..008M,2017MNRAS.472.1315W,2019PASP..131k8002M,2021arXiv211112142D}. 
This requirement poses a challenge given the CSST survey characteristics.
Empirical approaches can be further 
categorized into model-dependent \cite{2010ApJ...717...40K,2010A&A...514A..63P} 
and model-independent \cite{2007MNRAS.382..377W,2015MNRAS.451.1955W}
methods.
Model-independent methods primarily rely on analytical approaches 
that utilize multi-band colors. However, most CSST SNe do not have 
simultaneous color information.

We choose the model-dependent method of spectral energy distribution (SED) template fitting for 
CSST SNe classification, 
which is a physically motivated approach \cite{2008AJ....135..348S}. 
This method determines the SN subtype by minimizing the $\chi^2$ 
when comparing the observed magnitudes with the synthetic magnitudes 
computed from filter throughput curves and library templates.
The library templates used for the SED template-fitting method include
type Ia (Nugent-snIa, Nugent-sn91t, Nugent-sn91bg), type Ibc (Nugent-snIbc, Nugent-hyper), and
type II (Nugent-sn2p, Nugent-sn2l) from \cite{2002PASP..114..803N}.

We utilize the built-in function $mcmc\_lc$ in \textsc{Sncosmo}, 
employing the Markov Chain Monte Carlo method, 
to perform SED template fitting 
for the selected CSST SN samples.
Observations with SNR $< 5$ are considered too
weak and are excluded from the fitting process.
Each Nugent template includes five parameters to be fitted: $z,~t_0,~amplitude, mwebv$, and $hostebv$.

The CSST will capture images of the host galaxies
associated with the SNe.
According to \cite{Zhou_2022}, below redshift 1.4, 
the average photo-$z$ uncertainty of the host galaxy 
observed by the CSST is within 0.05. 
For galaxies with lower redshifts, the uncertainty is even smaller. 
Additionally, the redshift of the host galaxy can also be obtained from existing catalogs.
Therefore, during the fitting process, we 
apply a Gaussian prior distribution $\mathcal{N}(z_{true}$, 
$0.05\times(1+z_{true})$) to the $z$ parameter.

We assume that the parameter $mwebv$ is perfectly known 
from the dust map, and the $hostebv$ is fitted as
a variable parameter ranging from 0 to 0.5 mag.
The fitting function makes initial guesses for $t_0$ and $amplitude$ based on the data, then runs a minimizer.
We fit the simulated SN LCs with each of 
the seven templates in the library one by one. 
For each template, we obtain a $\chi^2$ value by fitting the data. 
After fitting all seven templates, we compare the $\chi^2$ values 
and select the template with the minimum $\chi^2$ as the one 
that best fits the data.

\begin{table}[H]
\footnotesize
\centering
\caption{Confusion matrix for binary classification}
\label{table: confusion matrix}
\begin{tabular*}{\columnwidth}{@{\extracolsep{\fill}}m{2cm}<{\centering}|c|c} 
    \hline \hline
     & \multicolumn{2}{c}{\textbf{Predicted label}}  \\ 
    \hline
    \textbf{True label} & Positive (P) & Negative (N)  \\ 
    Positive (P) & TP & FN \\ 
    Negative (N) & FP & TN \\
    \hline
\end{tabular*}
\end{table}

To evaluate the template fitting method 
performance, we use the confusion matrix to visualize
the good and bad classification. 
Table \ref{table: confusion matrix} shows the 
confusion matrix for binary classification.
True positive (TP) is the number of sources
predicted to be true and actual is true. False positive (FP) is
the number of sources predicted to be true and
actual is false. True negative (TN) is the number of sources 
predicted to be false and actual is false.
False negative (FN) is the number of sources predicted to be
false and actual is true. 
For ease of reading,
the confusion matrix is usually normalized
by dividing each entry by the true number of
each SN subtype.

Through the confusion matrix, 
the precision (also known as purity) and 
the recall (also known as completeness) 
are defined as
\begin{eqnarray}\label{eq:precision}
\rm precision = \frac{TP}{TP+FP},
\end{eqnarray}

\begin{eqnarray}\label{eq:recall}
\rm recall = \frac{TP}{TP+FN}.
\end{eqnarray}
The precision means the number of correct predictions in each class 
compared to the total number of predictions in that class, and
the recall means the number of correct predictions in each class 
compared to the total number of that class.

\section{CSST supernova mock samples} \label{sec:samples}
In this section, we make selection cuts to obtain a well-classified ``gold'' sample and an alert sample from the mock SNe data generated in the previous section. 
The results are summarized in Fig.~\ref{fig:redshift distribution of CSST SN} and Table~\ref{table:selection_cuts}.

\begin{table*}
\centering
\footnotesize
\begin{threeparttable}[left]
\caption{Number of SNe candidates after each selection cut}
\label{table:selection_cuts}
\begin{tabular*}{\textwidth}{@{\extracolsep{\fill}}lccccccc} 
    \hline \hline
     \multirow{2}{*}{Sample} & \multirow{2}{*}{Cut} & \multicolumn{3}{c}{Wide Field Survey} & \multicolumn{3}{c}{Deep Field Survey}\\
     \cmidrule(r){3-5} \cmidrule(r){6-8}
     & & Ia & Ibc & II & Ia & Ibc & II  \\
     \hline
     \multirow{2}{*}{Gold sample} & Q1 & \multicolumn{3}{c}{27,210 (Ia+Ibc+II)} &\multicolumn{3}{c}{4,546 (Ia+Ibc+II)}  \\
      & Q2 & 6,766 (91\%) & 1,943 (62\%) & 5,680 (92\%) & 675 (93\%) & 249 (66\%) & 773 (98\%)  \\
      \hline
      Alert sample & Q3  & 18,344 (61\%) & \rule[0.5ex]{2mm}{0.1mm} & 2,671 (49\%) & 1,365 (60\%) & \rule[0.5ex]{2mm}{0.1mm} & 207 (47\%)\\
    \hline
\end{tabular*}
\begin{tablenotes}
\item The numbers include contamination from other types of SNe, and the percentages in the parentheses are the classification precision, i.e., the percentage of correctly classified SNe.
\end{tablenotes}
\end{threeparttable}
\end{table*}

\begin{figure*}[t!]
\centering
\includegraphics[width= 2\columnwidth]{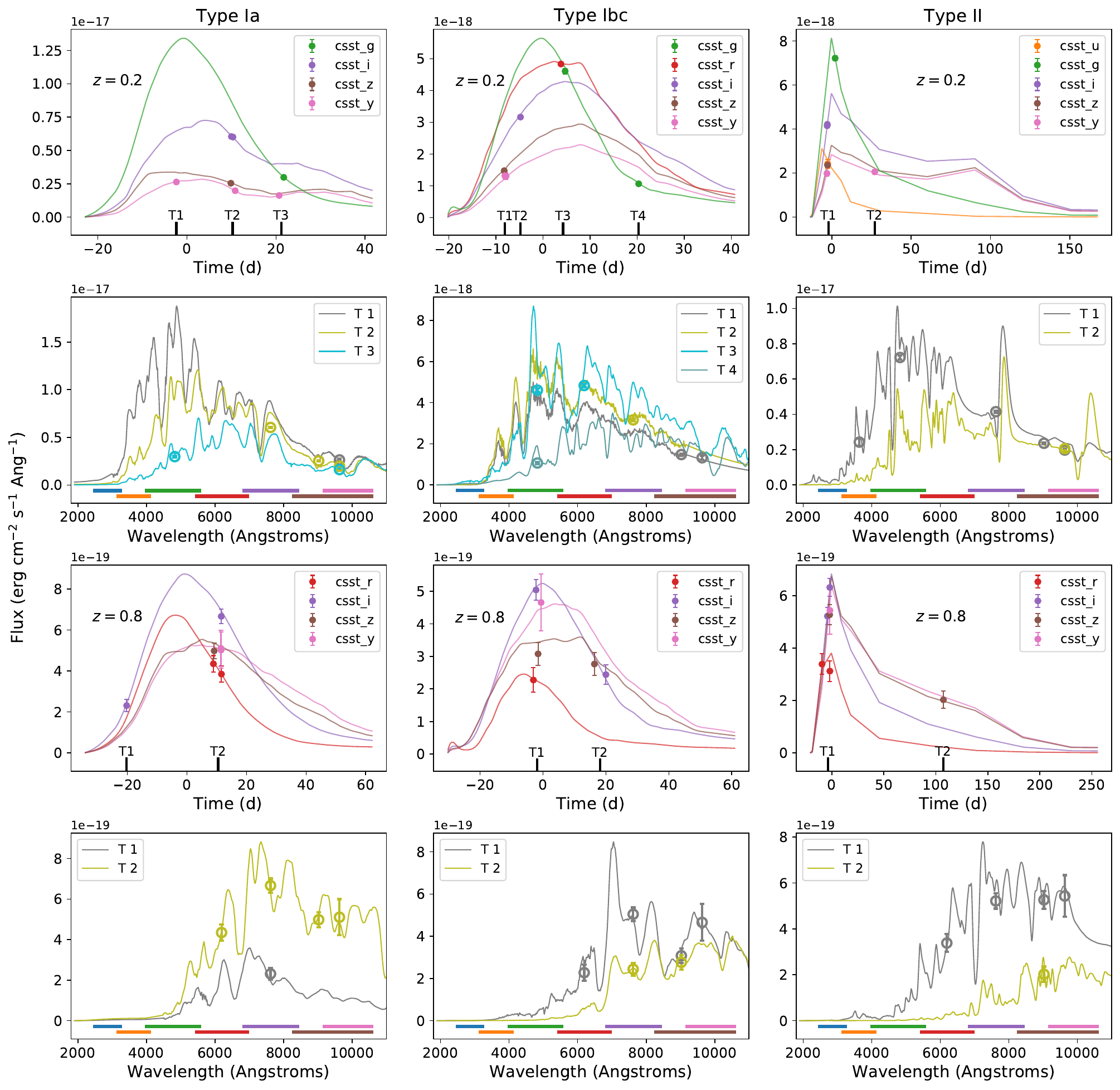}
\caption{
Examples of type Ia (left column), type Ibc (middle column), and type II (right column) SN mock observations along with the best-fit model LCs and SEDs. 
The top two rows are for the three selected SNe at $z=0.2$. The SEDs in each panel of the second row are plotted at the epochs (T$_1$, T$_2$, etc.) labeled in their corresponding first-row panel.
From left to right, the thick line segments in each panel of the second row mark the wavelength ranges of NUV, u, g, r, i, z, and y bands.
The bottom two rows are the same as the top two rows but for the three selected SNe at $z=0.8$.
    }
\label{fig:classification visualizations}
\end{figure*}

\begin{figure*}[t!]
\centering
\begin{subfigure}{}
\includegraphics[width=\columnwidth]{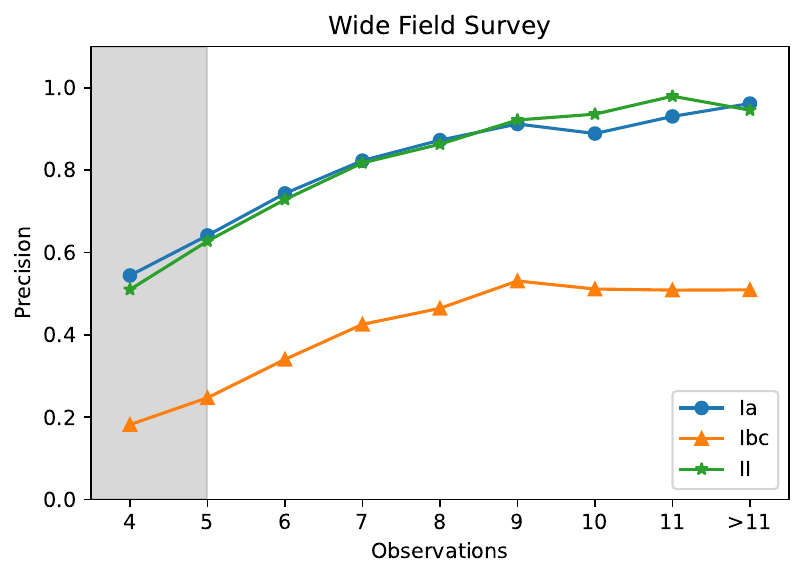}
\end{subfigure}
\begin{subfigure}{}
\includegraphics[width=\columnwidth]{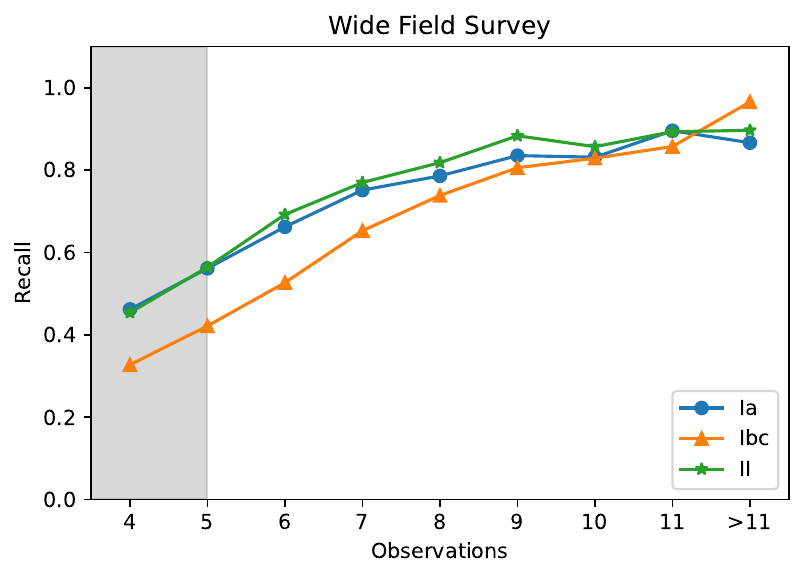}
\end{subfigure}
\begin{subfigure}{}
\includegraphics[width=\columnwidth]{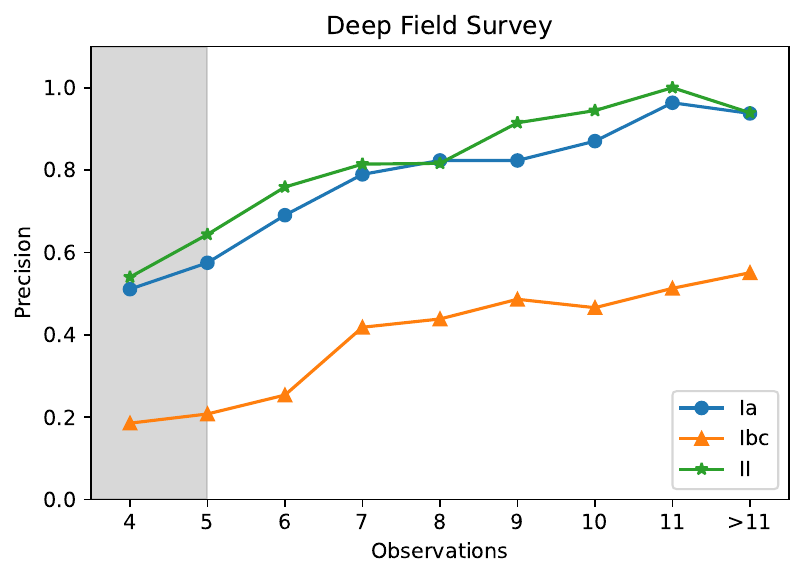}
\end{subfigure}
\begin{subfigure}{}
\includegraphics[width=\columnwidth]{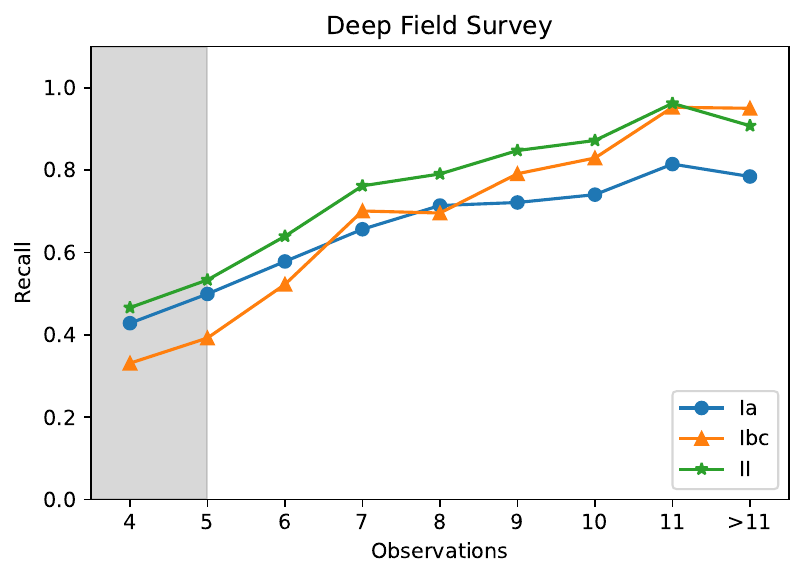}
\end{subfigure}

\caption{
The classification precision and recall versus the number of observations after the Q1 cut from the wide field survey (\textit{upper panels})
and the deep field survey (\textit{lower panels}).
SNe in the shaded region are discarded from further analyses.
}
\label{fig:classification of selected SN samples}
\end{figure*}

\begin{figure*}[t!]
\centering
\begin{subfigure}{}
\includegraphics[width=\columnwidth]{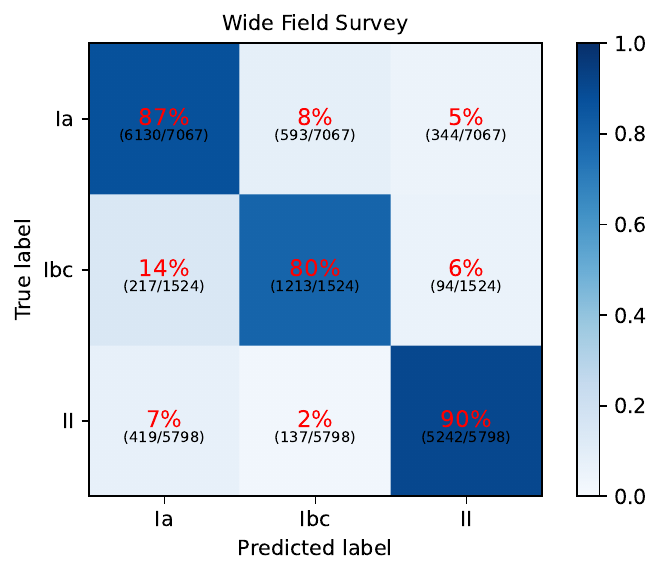}
\end{subfigure}
\begin{subfigure}{}
\includegraphics[width=\columnwidth]{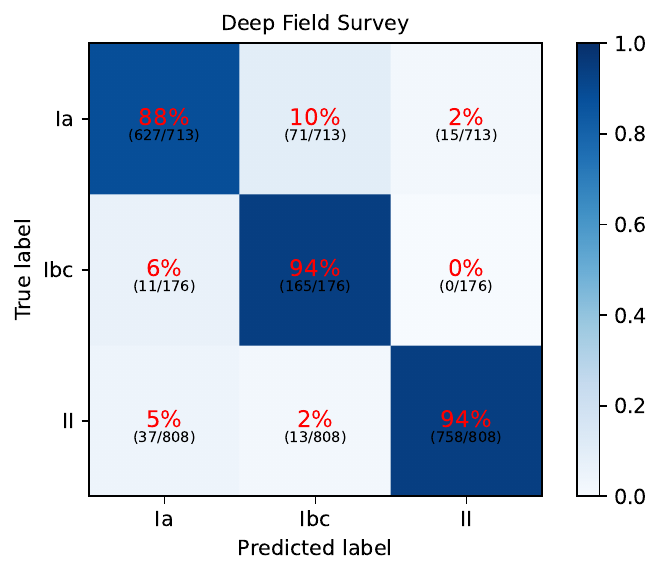}
\end{subfigure}
\caption{
The normalized confusion matrices of the gold sample (\textit{left panel:} the wide field; \textit{right panel:} the deep fields).
The horizontal axis represents the predicted SN subtypes, and the vertical axis represents the true SN subtypes. The numbers on the diagonal are the classification recall.
}
\label{fig:confusion matrix of golden sample}
\end{figure*}

\begin{figure*}[t!]
\centering
\begin{subfigure}{}
\includegraphics[width=2\columnwidth]{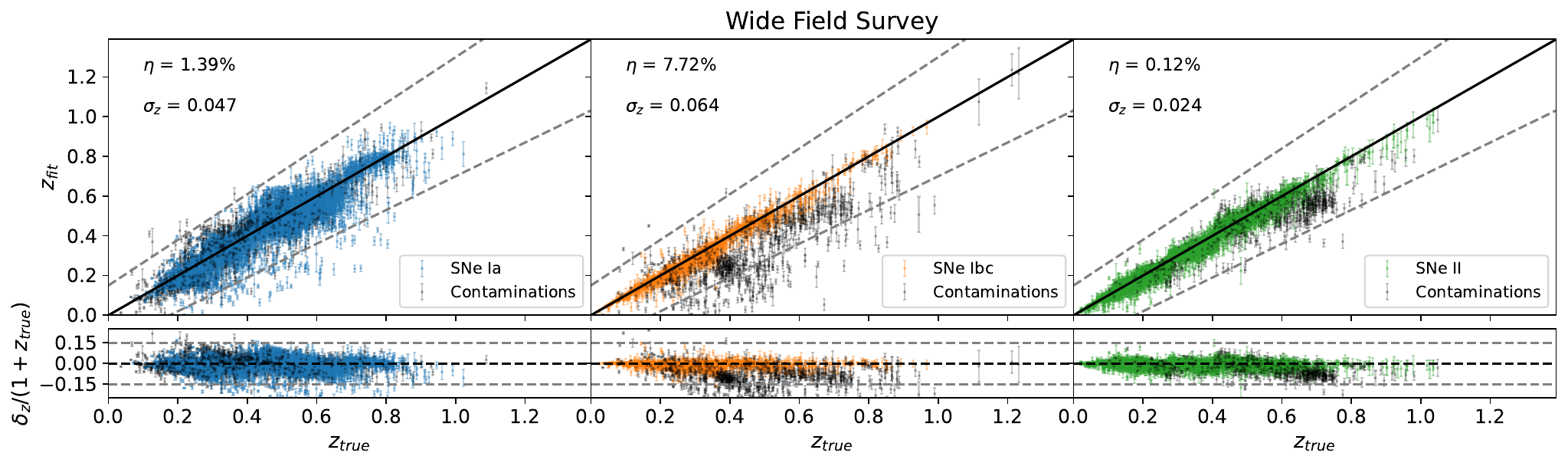} 
\end{subfigure}
\quad
\begin{subfigure}{}
\includegraphics[width=2\columnwidth]{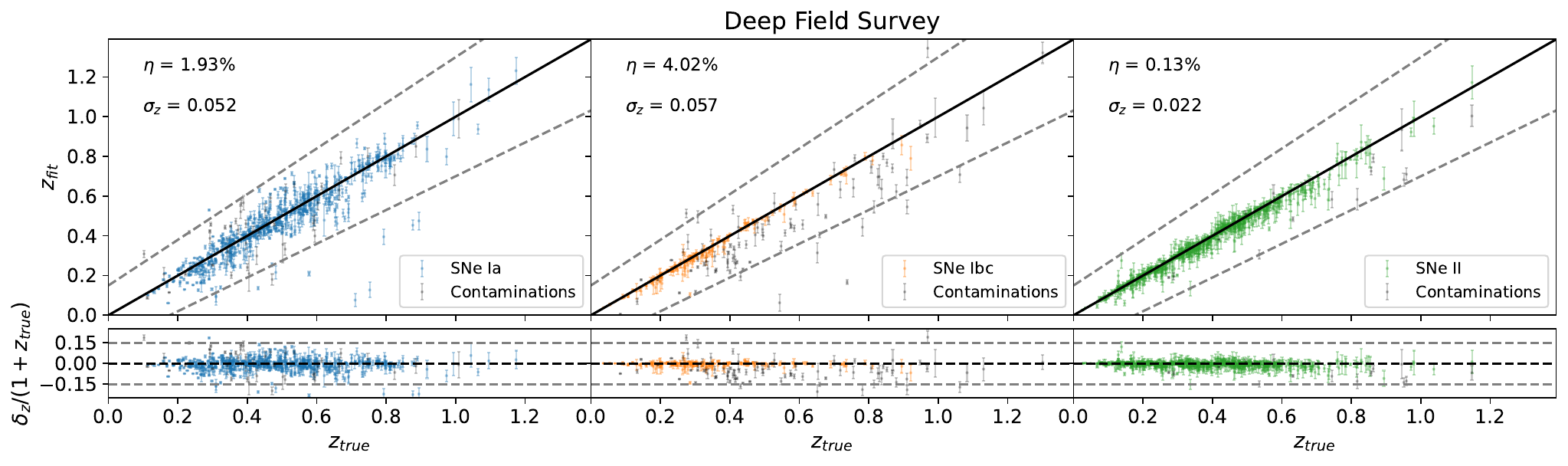}
\end{subfigure}
\caption{
The photo-$z$ estimation of the gold sample from 
the wide field survey (\textit{upper panels}) and the deep field survey (\textit{lower panels}). 
The $\sigma_z$ and $\eta$ represent the estimation uncertainty and the outlier fraction, respectively.
The gray points
represent the misclassified SNe.
    }
\label{fig:zfit_vs_ztrue of golden sample}
\end{figure*}

\subsection{Gold sample}
\label{well observed SNe classification}

Following the selection criteria outlined in
\cite{2012ApJ...753..152B,2018ApJ...867...23H,2023SCPMA..6629511L}, we implement a set of quality cuts, referred to as the Q1 cut, on the observed SNe before the fitting process:
\begin{enumerate}
  \setlength{\itemsep}{0pt}
  \item At least two observations in the same band,
        with one being a non-detection (SNR $<1$) and 
        the other having SNR $>5$.
  \item At least two different bands with SNR $>$ 5. 
  \item At least one observation with SNR $>$ 5 before the \textit{B}-band peak magnitude, and at least one observation with SNR $>$ 5 after the \textit{B}-band peak magnitude.
  \item At least six observations with SNR $>$ 5.
\end{enumerate}

As seen in Table~\ref{table:selection_cuts}, 27,210 SNe in the wide field and 4,546 SNe in the deep fields pass the Q1 cut. 
Fig.~\ref{fig:classification visualizations} provides several examples of the SN LCs and SEDs along with mock observations in the observer's frame.
Classification results of these SNe are shown in Fig.~\ref{fig:classification of selected SN samples}. 
For those relatively rare cases with a ``large'' number of observations, there is a good chance that several observations are made within a few days, providing little help on classification. Therefore, the precision and recall do not increase monotonically with the number of observations.

We further refine the sample with a Q2 cut that removes SNe with inadequate fits. The resulting sample is referred to as the gold sample, which may be used as a candidate catalog for constraining cosmology in future work. The fitting program returns the corresponding minimum $\chi^2$ values for each subtype (Ia, Ibc, and II), and we examine the relative $\chi^2$ value of the best-matching and second-matching subtypes based on \cite{2008AJ....135..348S}.
We denote the $\chi^2$ value of the best-matching subtype as $\chi^2_{min}$, the second-matching subtype as $\chi^2_{sec}$.
The Q2 cut is then described as follows:
\begin{enumerate}
\setlength{\itemsep}{0pt}
  \item The fit is convergent.
  \item $\chi^2_{min}<$ $\chi^2_{sec}-$ $\chi^2_{min}$.
  \item The reduced chi-square $\chi^2_{min}$/$n\_dof$ $<$ 10.
\end{enumerate}

The resulting gold sample, with the wide field and deep fields combined, contains about 7,400 SNe Ia, 2,200 SNe Ibc, and 6,500 SNe II candidates with overall classification precision of 91\%, 63\%, and 93\%, respectively. 
The redshift distributions of the subtypes are shown in Fig.~\ref{fig:redshift distribution of CSST SN}.
The recall rates for the sample's classification are provided in the normalized confusion matrices in Fig.~\ref{fig:confusion matrix of golden sample}. In the wide field, the recall rates for SN Ia, Ibc, and II are 87\%, 80\%, and 90\% respectively, while in the deep fields, they are 88\%, 94\%, and 94\% respectively.

Fig.~\ref{fig:zfit_vs_ztrue of golden sample} presents
the photo-$z$ estimation of the gold sample.
We define outliers to be $|\delta_z|/(1+z_{true}) > 0.15$, 
where $\delta_z = z_{fit} - z_{true}$. 
The standard deviation of $|\delta_z|/(1+z_{\text{true}})$ as $\sigma_z$ is also calculated to measure the uncertainty of the photometric redshift estimation, considering the redshift evolution.
In the case of SNe Ia, the photo-$z$ uncertainty is roughly the same as the somewhat conservative prior $0.05(1+z_{true})$ from the host galaxies. 
This means that improving the photo-$z$s of the host galaxies would directly improve those of the SNe Ia.
The errors of SNe Ibc themselves are relatively small, but they are affected by contamination from other types, leading to a degradation of the sample's photo-$z$s.
Since the Ibc subsample has a relatively low classification precision, it would not be very useful before further refinement. 
The SNe II subsample exhibits the best photo-$z$ performance despite the contamination.

\begin{figure*}[t!]
\centering
\begin{subfigure}{}
\includegraphics[width=\columnwidth]{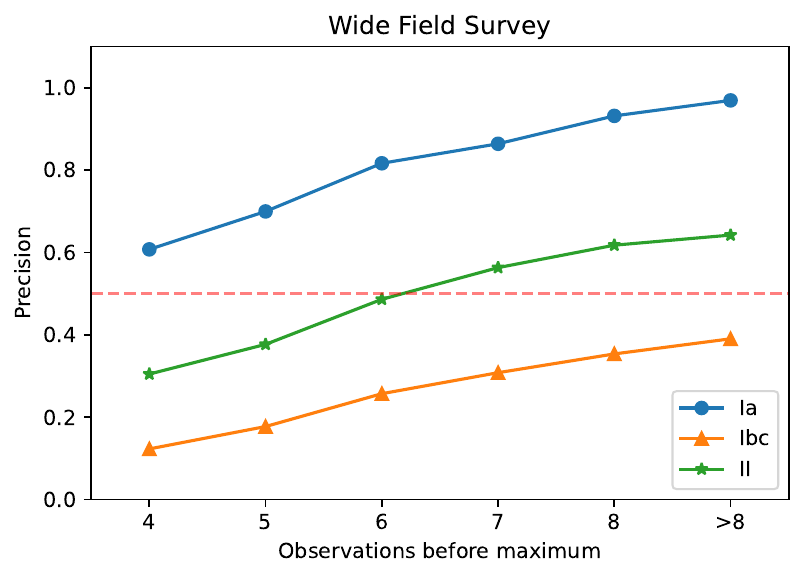}
\end{subfigure}
\begin{subfigure}{}
\includegraphics[width=\columnwidth]{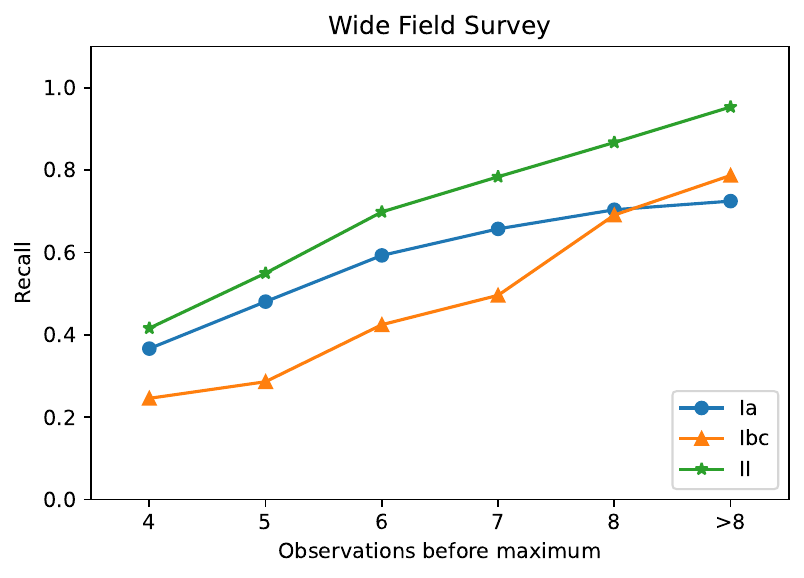}
\end{subfigure}
\begin{subfigure}{}
\includegraphics[width=\columnwidth]{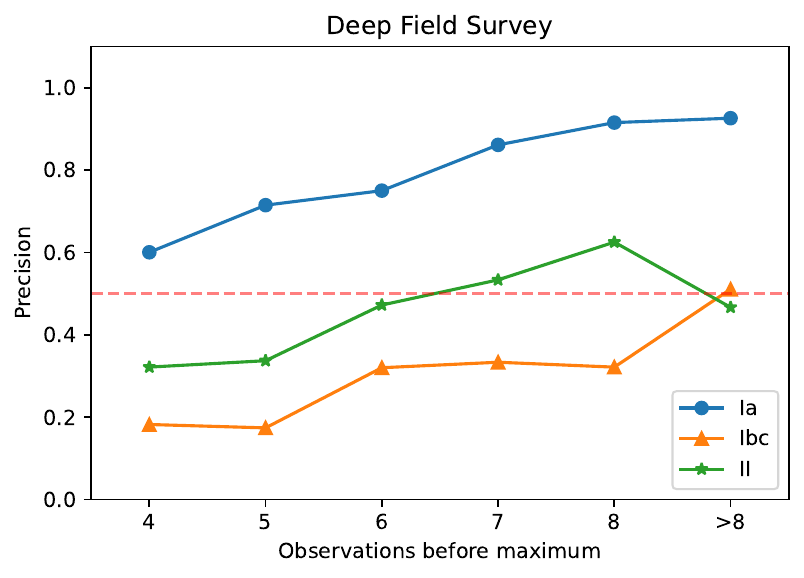}
\end{subfigure}
\begin{subfigure}{}
\includegraphics[width=\columnwidth]{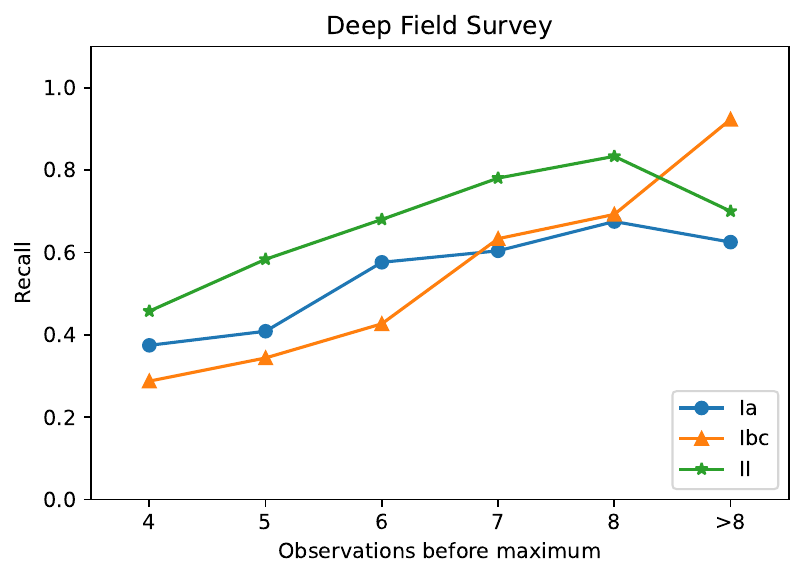}
\end{subfigure}
\caption{
The classification precision and recall versus the number of observations before the maximum for the pre-maximum sample. 
The red dash line marks the criterion of $\mathrm{precision}\gtrsim 50\%$.
}
\label{fig:alert classification of selected SN samples}
\end{figure*}

\subsection{Alert sample} \label{SN alerts}
Early detection of pre-maximum SNe is crucial for subsequent observations and studies. These include classifying candidates based on their spectra near the peak brightness, understanding how the spectra evolve from early to late phases, and accurately estimating SN model parameters through well-observed LCs.
The CSST's image quality and single-exposure depths help catch SNe during their early ascents.
Alerts could be issued at the earliest opportunity, facilitating prompt follow-up observations.

By requiring the SN to be observed in the same band at least twice before the \textit{B}-band peak luminosity, one with $\mathrm{SNR}<1$ and the other with $\mathrm{SNR}\ge 5$, we obtain a pre-maximum sample of approximately 680,000 SNe: about 340,000 SNe Ia, 110,000 SNe Ibc, and 190,000 SNe II in the wide field, and about 22,000 SNe Ia, 9,100 SNe Ibc, and 17,000 SNe II in the deep fields. 
Fig.~\ref{fig:redshift distribution of CSST SN} shows the redshift distributions of this sample. 
Most of these SNe would not receive enough CSST observations before the maximum to be correctly identified. 

One might want to set a minimum classification precision requirement to filter out an alert sample for further observations. However, in reality, the classification precision of the sample will not be known until proper follow-up observations have been conducted. Therefore, an indirect criterion is necessary.
As demonstrated in Fig.~\ref{fig:classification of selected SN samples}, the classification accuracy improves as the number of observations increases. One can therefore use the the number of observations before the maximum as a proxy for precision.
We consider $\sim 50\%$ to be an acceptable precision for a meaningful alert sample. Fig.~\ref{fig:alert classification of selected SN samples} translates it into $\ge 4$ and $\ge 6$ observations before the maximum for SNe Ia and SNe II, respectively.
These requirements and those for the pre-maximum sample are referred to as the Q3 cut collectively. 
The resulting alert sample, with the wide field and deep fields combined, contains about 20,000 SNe Ia (precision 61\%)
and 2,900 SNe II (precision 49\%) candidates.
SNe Ibc reach 50\% precision only in the deep fields with merely 47 candidates, so we do not include them in the alert sample.

\begin{figure}[H]
\centering
\includegraphics[width= \columnwidth]{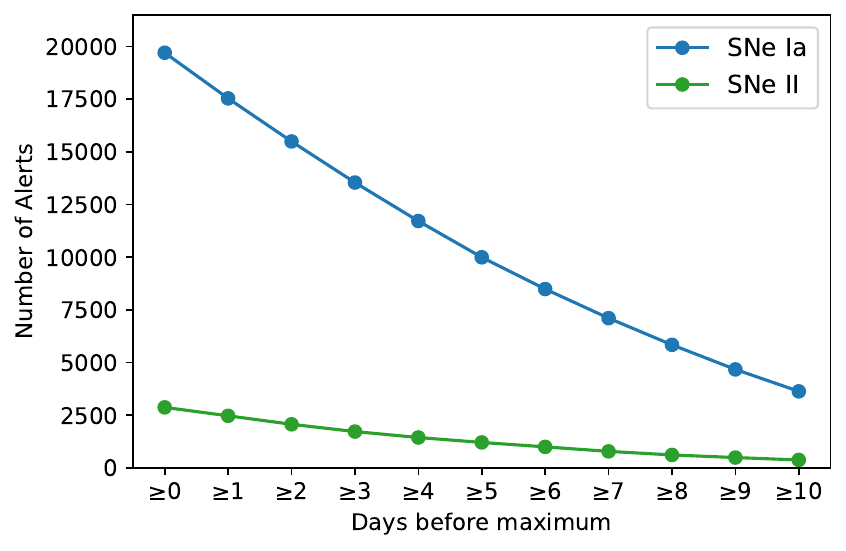}
\caption{
    Numbers of SN alerts that can be announced as a function of days before maximum light. 
    }
\label{fig:SNe alerts}
\end{figure}

Fig.~\ref{fig:SNe alerts} shows the number of alerts that can be issued as a function of days before maximum light. 
The lead time of follow-up observations can vary significantly from one facility to another. 
For a well-coordinated full-sky follow-up program that has a lead time of just two days, roughly 15,500 SNe Ia and 2,100 SNe II candidates are within its reach.
The numbers drop by 54\% and 62\% for SNe Ia and SNe II, respectively, if the lead time increases to seven days.

\section{Discussions and Conclusions} \label{sec:summary}

Our simulations show that the CSST wide field and deep fields can provide more than 16,000 well-classified SNe candidates of various types at $z\lesssim 1$ with classification precision above 90\% for type Ia and type II.
The proposed 9~deg$^2$ CSST ultra-deep field can contribute another $\sim 2000$ well-observed SN Ia at $z\lesssim 1.3$ for cosmology \cite{2023SCPMA..6629511L,2024MNRAS.530.4288W}.
Meanwhile, the CSST survey can trigger alerts for the detection of 
about 15,500 SNe Ia (precision 61\%)
and 2,100 SNe II (precision 49\%) candidates at least two days before maximum.
These samples will be of great value for SN science and cosmology. 

It is worth emphasizing the CSST's unique capability in the NUV. An example of important applications is searching and observing shock-cooling events. 
Early-time UV observations of CCSNe and the measurements of their optical LCs are crucial for comprehending the physics behind SN explosions and understanding the progenitor properties \cite{1978ApJ...223L.109K,2011ApJ...728...63R}. 
Only a handful of such events were detected 
in the past \cite{1989ARA&A..27..629A,2006Natur.442.1008C,2008ApJ...683L.131G,2010ApJ...720L..77G,2014Natur.509..471G,2022ApJ...924...55G}.
\cite{2008Sci...321..223S} proposed a method 
for identifying shock-cooling events: an optical survey dataset to locate the SN and a UV dataset to search for the associated shock-cooling event.

UV observations can capture CCSNe explosions in the very early phase
\cite{2024Natur.627..754L,2023SciBu..68.2548Z}.
We adopt the method in  Ref~\cite{2016ApJ...820...57G} to predict the
number of shock-cooling events of SNe II to be seen by the CSST, assuming that all the events are from red supergiant (RSG) progenitors with a single set of fiducial parameters. 
The peak absolute magnitude of the shock-cooling model \cite{2011ApJ...728...63R} in the NUV band is determined by the RSG parameters \cite{2016ApJ...820...57G}:
\begin{equation} \label{eq:Mpeak}
M_{\rm peak}^{\rm NUV} \approx -11.2 - 2.3\log_{10}\left(\frac{R_{\ast}}{R_{\odot}}\right) - 2.3\log_{10}\left(\frac{E}{10^{51}~\mathrm{erg}}\right),
\end{equation}
where the radius $R_{\ast}$ takes the fiducial value of $500~R_{\odot}$, and the energy $E=10^{51}$~erg normalized to ejecta mass of 10 $M_{\odot}$.
Using Eq.~(\ref{eq:Mpeak}), we obtain the peak absolute magnitude $M_{\rm peak}^{\rm NUV}\approx-17$ mag.
Given that the CSST wide-field survey reaches 24.5~mag in the NUV band with a single exposure (Table~\ref{table:multi-color imaging magnitude}), we expect it to be capable of detecting shock-cooling events up to $z\sim 0.3$.

The CSST survey camera has four detectors for NUV imaging, totaling a FoV of 0.14 $\rm deg^2$.
With the volumetric rate of CCSNe from Eq.~(\ref{eq:Cc rate}), we estimate that there are about three $z\le 0.3$ SNe II exploding within an area of the CSST NUV FoV every year.
We adopt one day as the early detection window for shock-cooling events \cite{2016ApJ...820...57G}. 

The probability of catching a shock-cooling event is then $3/365$, or 0.0082, per CSST pointing,
regardless of whether the pointing is fixed in the sky with a cadence $\ge 1$~day or tiles the sky without overlapping.

The CSST survey will complete about 65,000 pointings every year.
We ignore those repetitive observations within one day ($\sim 28\%$), 
which are helpful in general but do not significantly increase the probability of detecting shock-cooling events,  
and estimate the events observed every year to be $0.0082 \times 65,000 \times 0.72$, or 380. 
Therefore, the CSST can catch hundreds of shock-cooling events serendipitously every year in the NUV. 
This is comparable in numbers to the projection of the dedicated wide-field UV explorer ULTRASAT \cite{2014AJ....147...79S,2016ApJ...820...57G} but with a wider redshift span. 

Another unique aspect of the CSST is its slitless spectroscopy observations in the wavelength range 255-1000~nm. The slitless spectra will provide an additional dimension for in-depth studies on SNe. We will explore CSST spectroscopic SN samples in a future paper. 

Finally, the CSST will not be operating alone. Close collaboration with its contemporary projects such as the WFST, the Euclid mission, the Rubin Observatory, and the Roman Space Telescope will greatly enhance the science for all parties. 

\Acknowledgements{We would like to thank Subo Dong for helpful discussion.
This work was supported by the National Key R\&D Program of China No. 2022YFF0503400 and 2022YFF0503401, China Manned Space Program grant No. CMS-CSST-2021-B01, CMS-CSST-2021-B04, and CMS-CSST-2021-A12, Science Program of Beijing Academy of Science and Technology (24CD014), National Natural Science Foundation of China (NSFC grants 12288102 and 12033003), and Tencent Xplorer Prize.
This work made use of Astropy \cite{2013A&A...558A..33A,2018AJ....156..123A} and
Sncosmo \cite{2016ascl.soft11017B,2022zndo....592747B}.}

\InterestConflict{The authors declare that they have no conflict of interest.}




\end{multicols}
\end{document}